\documentclass{PoS}

\usepackage{latexsym}
\usepackage{amsmath}
\usepackage{epsfig}
\usepackage{graphics}
\title{Definition and parametrization of non-perturbative effects in quenched QCD}

\ShortTitle{non-perturbative effects in quenched QCD}

\author{A. Denbleyker, D. Du, \speaker{Y. Meurice}, and M. Naides\thanks{Undergraduate student at Cornell University, Ithaca NY; REU student at the University of Iowa in 2006.}\\
Department of Physics and Astronomy\\ The University of Iowa\\
Iowa City, Iowa 52242, USA\\

        E-mail: \email{yannick-meurice@uiowa.edu}}
 
\abstract{
The notion of a non-perturbative effect is ambiguous if it requires the 
subtraction of a perturbative part defined by a diverging series.
A common procedure consists in dropping  
the order of minimal contribution and the higher orders.
This allows us to isolate very accurately the one-instanton effect 
for the double-well potential. 
For the one plaquette gauge theory, an exact analytical expression can be written for the non-perturbative part.  
We report recent attempts to extend this approach to the 
average plaquette of quenched QCD. Our goal is to express the non-perturbative effects in terms of expressions of the form $(\beta)^B {\rm e}^{-A\beta}$ 
calculable semi-classically. The situation is complicated by zeroes of the partition function in the complex $\beta$ plane (presumably near $5.75\pm 
i 0.2$). We discuss two methods to describe the intermediate and large order behavior of the 
perturbative series. One is inspired by mean field theory (logarithmic specific heat) and reproduces accurately the known perturbative series with only two free parameters.  
A diagrammatic interpretation of this fact is still lacking. 
The other is based on infra-red renormalons with a factorial growth showing up at order larger than 20 and a possible effective theory interpretation. 
These extrapolations are compatible with the 
non-perturbative part of the plaquette being proportional to $a^4$. We propose an exponential parametrization to the corrections to the universal part of the beta function and find results compatible 
with the suggestion of $a^2$ corrections made by C. Allton. }

\dedicated{To the memory of Thomas Branson and James Van Allen}

\FullConference{XXIV International Symposium on Lattice Field Theory\\
		 July 23-28 2006\\
		 Tucson Arizona, US}

\begin{document}
In these proceedings, we will discuss the 
definition of  the nonperturbative part of the average plaquette and its parametrization in terms of expressions of the form 
$A \beta^B{\rm e}^{-C\beta}$. There is a reasonable case that this is feasible for Wilson action in the fundamental and the hope is that 
$A$, $B$ and $C$ could be calculable 
semi-classically. Technical details regarding recent progress are available in a recent preprint \cite{npp}. Instead of trying to cover most of the material discussed in this preprint, 
we will rather emphasize selected points that we have disccused with some of the participants during the conference. 

We would like to emphasize that the commonly used  ``rule of thumb'' for perturbative series (at a given coupling, you drop the order in perturbation theory that gives the smallest contribution and all the higher orders), is a poor man substitute for 
regularizing the diverging perturbative series by introducing a large field cutoff or a large 
action cutoff as recently proposed \cite{convpert,effects04}. 
The field cutoff can then be fixed by an optimization procedure based on the strong coupling expansion \cite{optim,plaquette}. However, a first step consists in understanding the error associated with this simple procedure.
For a generic asymptotic series $A\sim \sum_k a_k \beta^{-k}$, we can define the 
error at order $k$:
\begin{equation} \Delta _k (\beta) \equiv A_{numerical}(\beta) - \sum_{l=0}^{k} a_l\beta ^{-l}\ .\end{equation}
If we assume that 
$\Delta_k \ \simeq \beta ^{-k-1} a_{k+1}$ ,
 (for $\beta$ large enough)
and that the 
large order behavior is $|a_k|\sim |C_1||C_2|^k\Gamma(k+C_3)$,  
we find \cite{npp} that the minimal error is 
\begin{equation}
Min_k \ |\Delta_k|\simeq \sqrt{2\pi}|C_1|( |C_2|/\beta)^{1/2-C_3}
{\rm e}^{-\frac{\beta}{|C_2|}}\ . 
\end{equation}
This equation works well for the ground state of the anharmonic oscillator where the exponential suppression 
can be interpreted semi-classically. For the double-well, it only works if we take the 
average of the two lowest energy states in order to cancel the one-instanton effect.
The one-instanton is a non-perturbative effect larger than the error estimated from the 
the large order behavior of the perturbative series. 

The separation between the perturbative and the non-perturbative part is a computational 
commodity. There is no claim that this separation has a physical meaning, unless one can find 
observables that are zero in perturbation theory, for example the difference between the two lowest energy states for the double-well potential. A simple example \cite{plaquette} where one can see exactly what is in the non-perturbative part is the $SU(2)$ one plaquette model 
\begin{equation}
Z(\beta)=\int_{SU(2)} dU {\rm e} ^{-\beta(1-\frac{1}{2}Re TrU)} =Z_{Pert.}(\beta)+Z_{NPert.}(\beta)\ .
\end{equation}
Due to the compactness of $SU(2)$, there is no large field problem. 
It is possible to expand $Z$ as a converging sum of $\beta$-dependent coefficients, denoted 
$A_k(2\beta)$ below,  
multiplying $\beta^{-k}$. We can get the regular perturbative series by decompactifying the integrals, namely adding the tails of integration from $2\beta$ to $\infty$, in order to get $\beta$-independent coefficients. 
At first sight this amounts to neglecting ${\rm e}^{-2\beta}$ effects, however, this affects the large order of the perturbative coefficients, which now grow at a factorial rate. 
In order to minimize the perturbative error, we truncate the series at order ${r(k^\star)}$ the integer closest to the one obtained with the procedure discussed above (see Ref. \cite{npp,plaquette} for details). 
More explicitly, 
\begin{equation}
Z_{Pert.}(\beta)=(\beta\pi)^{-3/2} 2^{1/2} 
	\sum_{k=0}^{r(k^\star)} A_k(\infty)\beta^{-k}\ , 
\end{equation}
with 
\begin{equation}
	A_k(x)\equiv 2^{-k}
	\frac{\Gamma(k+1/2)}{k!(1/2-k)}\int_0^{x}dt {\rm e}^{-t}t^{k+1/2}\ .
	\label{eq:al}
\end{equation}
The nonperturbative part consists in the dropped terms minus the added tails: 
\begin{equation}
Z_{NPert.}(\beta)=(R(\beta)-T(\beta))\ ,
\end{equation}
\begin{equation}
R(\beta)=(\beta\pi)^{-3/2} 2^{1/2}\sum_{r(k^\star)+1}^{\infty} A_k(2\beta)\beta^{-k}\ ,
\end{equation}
\begin{equation}
T(\beta)=(\beta\pi)^{-3/2} 2^{1/2}
 \times  \sum_{k=0}^{r(k^\star)}\beta^{-k} 
\frac{\Gamma(k+1/2)}{k!(1/2-k)}
\int_{2\beta}^{\infty}dt {\rm e}^{-t}t^{k+1/2} \ .
\end{equation}
A detailed calculation shows that at leading order, $R$ and $T$ are both proportional to 
$\beta^{-3/2}{\rm e}^{-2\beta}$, while their difference is proportional to $\beta^{-2}{\rm e}^{-2\beta}$ as predicted by the asymptotic behavior of the regular perturbative series. One can improve the accuracy of the result by keeping a finite range of integration for the coefficients and optimizing the range using the strong coupling expansion. If the order is large enough, it essentially amounts to use the bound $(t\leq 2\beta)$ provided by the compactness of the group \cite{plaquette}. 

The same program can in principle be carried on for quenched $QCD$ in 4 dimensions, however, keeping track of multiple tails of integration is a non-trivial task. 
We believe that the best approach is to use the stochastic perturbation method \cite{direnzo2000} with boundaries or periodic boundary conditions in configuration space. 
Before attempting this calculation for quenched QCD, we have considered models where independent numerical calculations are possible, namely one variable integrals and ground states of simple quantum mechanical systems \cite{tractable,asymp}. In absence of boundary, 
stochastic perturbation theory works quite well for these models, however, introducing a boundary is a delicate procedure in part because results may depend on few configurations. M. Naides 
has investigated three methods, reflection at the boundary, rejection of updates going out of the boundary and a-posteriori elimination of configurations outside of the boundary and 
concluded in favor of the third method. We are planning to compare these methods for quenched QCD. 

In the following, we will restrict the discussion to the accuracy of the regular perturbative series of the average plaquette in quenched QCD ($N=3$, $D=4$)
\begin{equation}
P(\beta)=(1/\mathcal{N}_p)\left\langle \sum_p
(1-(1/N)Re Tr(U_p))\right\rangle \ .
\end{equation}
We used a series in $\beta^{-1}$ calculated \cite{direnzo2000} up to order 10. 
A figure with coefficients up to order 16 is also available \cite{rakow05}. 
Extrapolated estimators \cite{third} indicate that 
$P\propto (1/5.74-1/\beta)^{1.08}$ in good agreement with a ratio analysis \cite{rakow2002}. This type of behavior is  
not expected. The power series in $\beta^{-1}$ should have a zero radius of convergence because the plaquette 
changes discontinuously at $\beta \rightarrow \pm \infty$ \cite{gluodyn04}. 
In addition, a singularity exactly on the real axis is not 
seen in 2d derivative \cite{third} of $P$. Such a singularity would requires massless glueballs which are not known to exist for the pure fundamental action. 
A plausible explanation \cite{third} is that the singularity is slightly off the real axis. 
If we assume an approximate logarithmic behavior for the specific heat as in mean field theory 
\begin{equation}
-\partial P/\partial \beta \propto {\rm ln}((1/\beta_m-1/\beta)^2+\Gamma^2)\ ,
\label{eq:mf}
\end{equation}
we obtain peaks and series compatible with what we know provided that $5.7 <\beta_m <5.9$ and 
$0.001<\Gamma <0.01$. This suggests the existence of zeroes of the partition function in the complex $\beta$ plane with 
\begin{equation}
0.03\simeq0.001\beta_m^2<|{\rm Im} \beta| <0.01\beta_m^2\simeq 0.33
\end{equation}

A. Denbleyker has performed direct searches for such zeroes using reweigthing methods. 
On a $4^4$ lattice, he found a pair of zeroes near $\beta=5.55\pm i0.1$, clearly within the 
region of confidence and in good agreement with the results on $4L^3$ lattices of Ref. \cite{alves}. We are not aware of existing results on larger isotropic lattices. 
On a $8^4$, we found no clear evidence for zeroes with $|{\rm Im} \beta| <0.03$. Our preliminary findings are shown in Fig. 1. The radii of confidence have been 
drawn without taking into account the autocorrelations which should slightly reduce these radii. Unlike the $4^4$ case, there are no zeroes clearly within the respective radii
of confidence. As the radius of confidence increases like the square root of the logarithm of the number of independent configurations, it will be difficult to find direct evidence for 
zeroes with $|{\rm Im} \beta| \approx 0.1$. Consequently, it would be useful to introduce a positive adjoint term in the action in order to move the zero closer to the real axis. If this works, we could follow the zero as $\beta_{adjoint}$ is decreased and try to extrapolate to $\beta_{adjoint}=0$. 
\begin{figure}
\begin{center}
\includegraphics[width=0.8\textwidth]{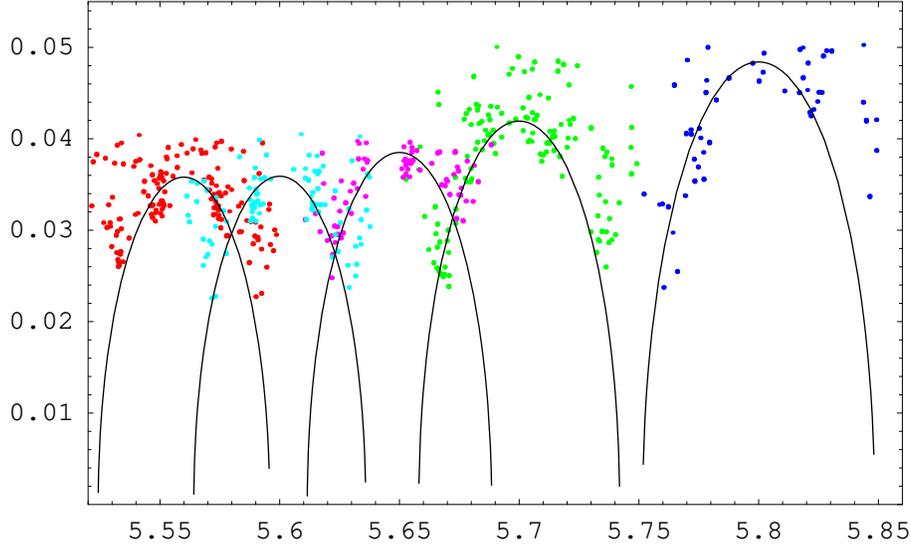}
\label{fig:zeroes}
\caption{Zeroes of the partition function in the complex $\beta$ plane  for a $8^4$ lattice 
using reweightings at $\beta_0$ = 5.55 (red), 5.6 (cyan), 5.65 (magenta), 5.7 (green) and 5.8 (blue). The dots correspond to distinct bootstraps and the solid lines to the successive radius of confidence }
\end{center}
\end{figure}

A pair of complex singularities near $\beta = 5.8$ is about all we need to account for most 
of the known terms of the perturbative series. 
Integrating the mean field ansatz of Eq. (\ref{eq:mf}), we obtain 
\begin{equation}
P_{pert.}= \sum _{k=0} b_k\beta^{-k}\simeq K({\rm Li}_2 (\beta^{-1}/(\beta_m^{-1}+i\Gamma))+{\rm h.c}\ ,
\label{eq:dilog}
\end{equation}
with ${\rm Li}_2(x)=\sum _{k=0}x^k/k^2 \ .$ 
We fixed $\Gamma=0.003$ and obtained $K=0.0654$ and $\beta_m$=5.787 using the known values 
of $a_9$ and $a_{10}$. As shown in Ref. \cite{npp}, this simple model agrees well with the 
numerical values from order 3 to 16. 
This fact should have a simple 
Feynman diagram interpretation.

A pair of complex singularities cannot govern the asymptotic behavior of a series that must have a zero radius of convergence. At this point, the best we can do is to use what is known in the continuum about the Borel transform of the series \cite{mueller93,itep} and assume that it 
can be connected to the lattice \cite{burgio97} by a coupling redefinition $\bar{\beta}=\beta(1+d_1/\beta+\dots)$:
\begin{equation}
P_{pert.}=\sum _{k=0} \bar{b}_k\bar{\beta}^{-k} \propto \int_{t_1}^{t_2}dt {\rm e}^{-\bar{\beta}t}\ (1-t\ 33/16\pi^2)^{-1-204/121}\ .
\label{eq:ir}
\end{equation}
Note that $t_1=0$ corresponds to the UV cutoff, 
$t_2=16\pi^2/33$ to the Landau pole. 
In the continuum, it has been argued \cite{itep} that it is necessary to introduce the 
gluon condensate in order to keep $t_2$ low enough and regularize the perturbative 
series, exactly as we advocated in lattice gauge theory. 
In the following, we take
$t_2 = \infty$ in order to get a regular perturbative series (using the one-plaquette analogy). 
This model has a minimal perturbative error of the form 
\begin{equation}
Min_k \ |\Delta _k|\simeq 3.5(\bar{\beta})^{204/121-1/2}{\rm e}^{-(16\pi^2/33)\bar{\beta}} \ .
\label{eq:guessintmodel}
\end{equation}
Except for the -1/2 in the exponent this is the two loop RG invariant. This is reminiscent of 
the $\beta^{-1/2}$ found in the one plaquette case. 
Following this analogy further, Eq. (\ref{eq:guessintmodel}) could in principle be compared with
what would be obtained from the probability distribution for one plaquette after integrating over all the other links.

The two models defined by Eqs. (\ref{eq:dilog}) and (\ref{eq:ir}) yield similar coefficients up to order 20-25. Above order 25, the renormalon model has larger coefficients growing at a factorial rate. D. Du has been trying to 
put together the two types of behavior in the framework of dispersion relations.
These two models are compatible \cite{npp} with the idea already put forward \cite{rakow2002,rakow05} that the non-perturbative part of the plaquette 
scales like $a^4$ with $a(\beta)$ defined with the force method \cite{force01, force98}. 

\begin{figure}[t]
\includegraphics[width=2.8in]{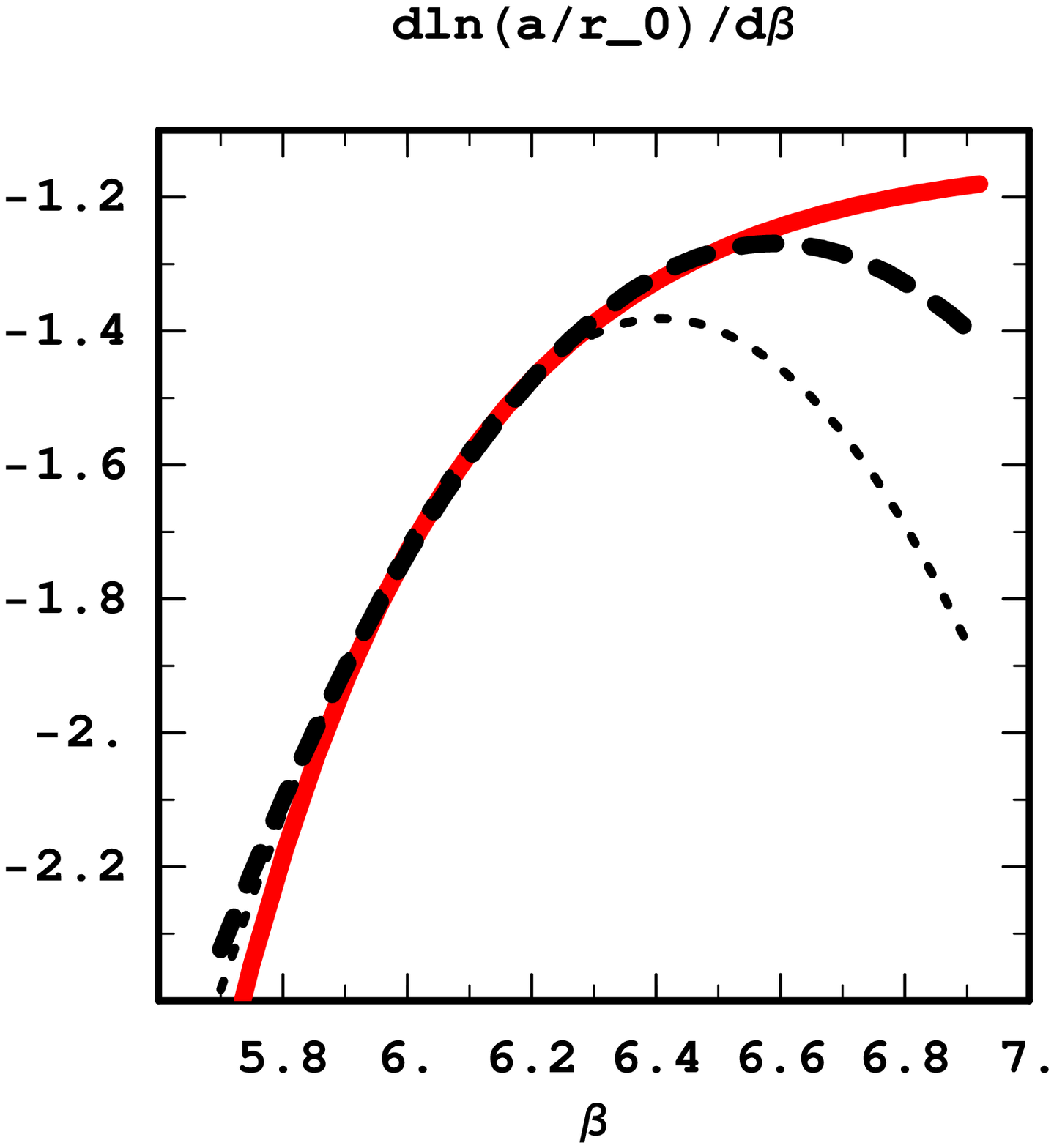}
\includegraphics[width=2.8in]{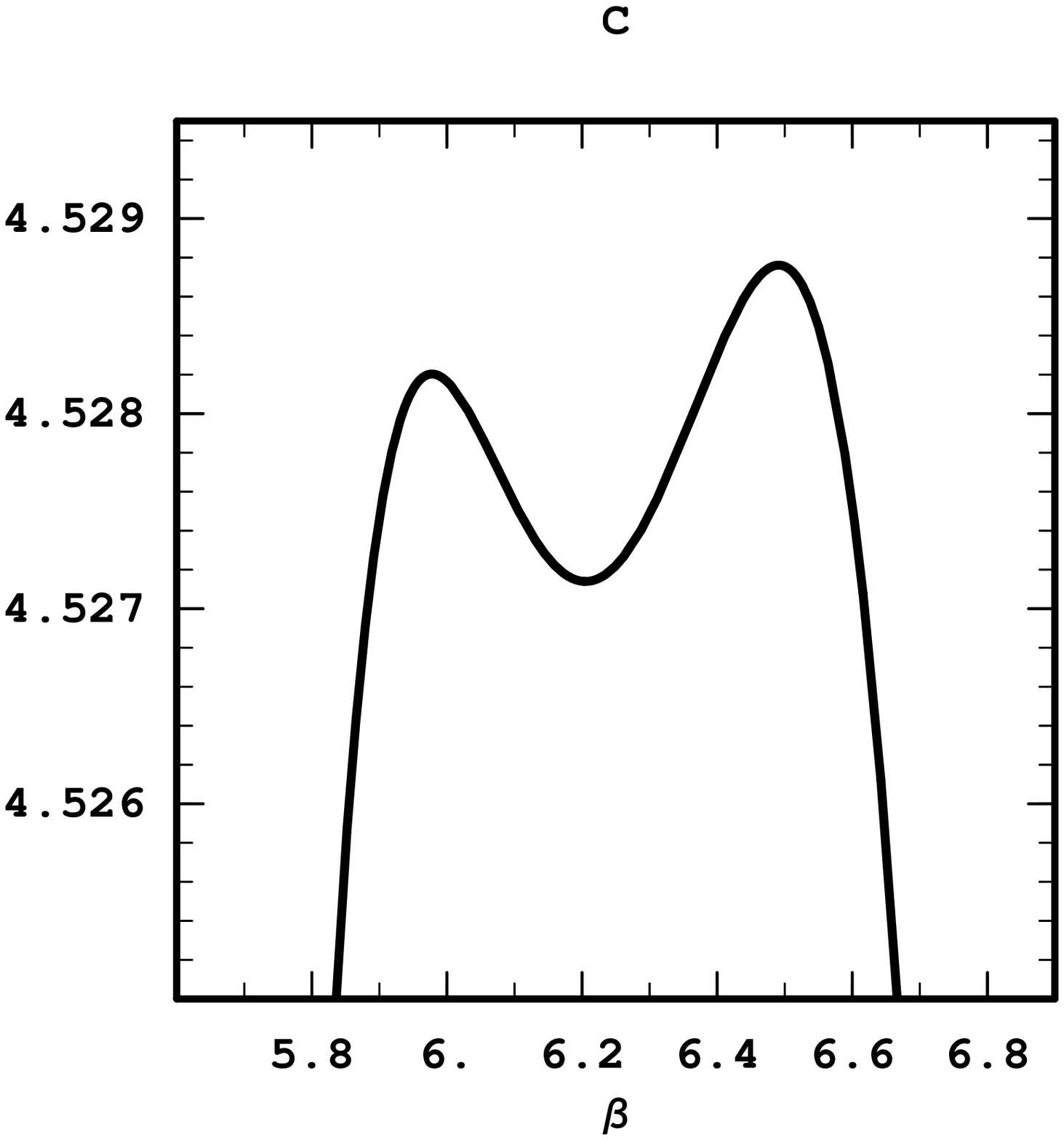}
\caption{Left: $ d {\rm ln}(a/r_0)/d\beta$ using Ref.  \cite{force01}  (thick dashes), Ref. \cite{force98}  (small dashes) and our parametrization (solid red line).
Right: estimation of the (approximate) constant $C$ in Eq. (16).}
\label{fig:deforce}
\end{figure}

The question is now: can we parametrize $a(\beta)$ in a way that would be suggestive from a semi-classical point of view? For this purpose, we start with a Taylor expansion \cite{force01} of $a(\beta)$ expressed in units of $r_0=0.5$ f in the interval $5.7<\beta<6.92 $:
\begin{equation}
{\rm ln}(a/r_0)= -1.6804 - 1.7331\ (\beta - 6) +0.7849\ (\beta - 6)^2 - 
      0.4428\ (\beta - 6)^3  \ .
\label{eq:force}
\end{equation}
We propose the parametrization \cite{npp}
\begin{equation}
d {\rm ln}(a/r_0)/d\beta=-(4\pi^2/33) + (51/121)\beta^{-1}+A{\rm e}^{-B\beta} \ .
\label{eq:fitforce}
\end{equation}
As shown in Fig. \ref{fig:deforce}, in the interval $5.9<\beta<6.4 $, the  derivative of Eq. (\ref{eq:force}) can be fit using $A=-1.35 \  10^{7}$ and $B=2.82$. It should be noted that when 
$\beta> 6.4$, there is a significant difference between the derivative of Eq. (\ref{eq:force}) and the derivative of a similar expression provided in a previous paper \cite{force98}.
The difference is hardly visible if we plot ${\rm ln}(a/r_0)$ in the interval $5.6<\beta<6.6  $, but when we take the derivative, the large constant term disappears and the cubic term 
that varies the most between the two expressions is amplified by a factor 3. This explains the large differences observed in Fig. \ref{fig:deforce} when $\beta> 6.4$. It would be interesting to try to get more accurate Taylor expansions in this specific region. 

We can now integrate and fit the constant of integration denoted $C$:  
\begin{equation}
{\rm ln}(a/r_0)=C-(4\pi^2/33)\beta + (51/121){\rm ln}(\beta)-(A/B){\rm e}^{-B\beta}\ .
\label{eq:ln}
\end{equation}
Fig. \ref{fig:deforce} shows that  
 $C=4.528(1)$ provides a good fit in the region where we have good agreement with both Refs. \cite{force01,force98}.  A nice plateau appears between $\beta = 5.9$ and 6.3. The extremal values in this interval are 4.5272 and 4.5282. It is also possible to obtain the lattice  scale $\Lambda_L$ from the constant of integration 
$C$, namely $\Lambda_L={\rm exp}(-C)/r_0\simeq 4.4\ MeV$. 
It is possible to use Eq. (\ref{eq:ln}) to predict ${\rm ln}(a/r_0)$ at large $\beta$. 
for instance, at $\beta=7.5$, we obtain -3.59 (for -3.63 in Ref. \cite{guagnelli02}) and 
-4.74 at $\beta=8.5$ (for -4.81 in Ref. \cite{guagnelli02})

Note that the value of $B$ seems consistent with the idea \cite{allton} 
of using $a_{pert.}^2$ corrections for this quantity. The symbol $a_{pert}$ refers to the 
one-loop or two-loop expression which in the short $\beta$ interval considered here 
can hardly be distinguished from each other.
The assumption of $a_{pert.}^2$ corrections fixes $B=8\pi^2/33\simeq 2.4$ which is close to the value 2.82 obtained above. 

There remain many challenges: check the exponential parametrizations more accurately at large $\beta$, calculate the parameters determined numerically semi-classically, find the precise 
connection between the non-perturbative part of the plaquette and what is called the gluon condensate in the continuum and find improved perturbative methods with a smaller error in the crossover region. 

We acknowledge valuable discussions at the conference with many participants and in particular C. Allton, M. Creutz, F. di Renzo, A. Duncan and A. Velytsky. 
This 
research was supported in part  by the Department of Energy
under Contract No. FG02-91ER40664. A. Denbleyker was supported by a Van Allen research grant. 
M. Naides was supported by a REU grant NSF PHY-0353509.

\end{document}